\documentclass[sigconf]{acmart}

\usepackage{xspace}
\usepackage{textcomp}
\usepackage{soul}
\usepackage{pifont}
\usepackage{marginnote}
\usepackage{hyperref}
\usepackage{amsfonts}
\usepackage{graphicx}
\usepackage{float}
\usepackage{subfig,rotate}
\usepackage{verbatim}
\usepackage{comment}
\usepackage{versions}
\usepackage{multirow}
\usepackage{longtable}
\usepackage{booktabs}
\usepackage{makecell}
\usepackage{caption}
\usepackage{acro}
\usepackage{tabularx}
\usepackage{adjustbox}
\usepackage{listings}
\usepackage{color}

\newfloat{lstfloat}{htbp}{lop}
\floatname{lstfloat}{Listing}

\def\bitcoin{%
  \leavevmode
  \vtop{\offinterlineskip 
    \setbox0=\hbox{B}%
    \setbox2=\hbox to\wd0{\hfil\hskip-.03em
    \vrule height .3ex width .15ex\hskip .08em
    \vrule height .3ex width .15ex\hfil}
    \vbox{\copy2\box0}\box2}}

\definecolor{dkgreen}{rgb}{0,0.6,0}
\definecolor{gray}{rgb}{0.5,0.5,0.5}
\definecolor{mauve}{rgb}{0.58,0,0.82}

\acmConference[Technical Report]{}{July}{2019}
\setcopyright{none}
\renewcommand\footnotetextcopyrightpermission[1]{}

\begin{document}

\title[Characterizing Bitcoin donations to open source software on GitHub]{\huge Characterizing Bitcoin donations to open source software on GitHub}

\author{Yury Zhauniarovich}
\authornote{Research was conducted while at Qatar Computing Research Institute, HBKU.}
\affiliation{
  \institution{Perfect Equanimity}
}
\author{Yazan Boshmaf}
\affiliation{
  \institution{Qatar Computing Research Institute, HBKU}
}
\author{Husam Al Jawaheri}
\authornotemark[1]
\affiliation{
  \institution{University of Luxembourg}
}
\author{Mashael Al Sabah}
\affiliation{
  \institution{Qatar Computing Research Institute, HBKU}
}
\renewcommand{\shortauthors}{Zhauniarovich et al.}

\begin{abstract}

Web-based hosting services for version control, such as GitHub, have made it easier for people to develop, share, and donate money to software repositories. In this paper, we study the use of Bitcoin to make donations to open source repositories on GitHub. In particular,  we analyze the amount and volume of donations over time, in addition to its relationship to the age and popularity of a repository.

We scanned over three million repositories looking for donation addresses. We then extracted and analyzed their transactions from Bitcoin's public blockchain. Overall, we found a limited adoption of Bitcoin as a payment method for receiving donations, with nearly 44 thousand deposits adding up to only 8.3 million dollars in the last 10 years. We also found weak positive correlation between the amount of donations in dollars and the popularity of a repository, with highest correlation ($r$=0.013) associated with number of forks.

\end{abstract}




\maketitle

\section{Introduction}
\label{sec:introduction}

Open source software has revolutionized our world. Developers around the world can now explore open source repositories, learn best practices, and collaborate on improving software used by millions of users. Web-based hosting services for version control, such as GitHub~\cite{GitHub}, GitLab~\cite{GitLab}, and Bitbucket~\cite{Bitbucket}, have played an important role in this revolution. Developers today use these services to facilitate all stages of software engineering, including planning, version control, issue tracking, release management, and software deployment. Among these platforms, GitHub stands out as the most popular with more than 96 million repositories~\cite{octoverse}, out of which more than three million are open sourced under MIT, Apache, GPL, or other open source software licenses~\cite{GitHubActivityDataset}. 

As open source repositories require considerable time and money investments by developers, it is customary to ask users for financial aid, often in the form of donations, to support and maintain their development. While there are several platforms that facilitate this process~\cite{OpenSourceFundingPlatforms_Kumar2018}, they typically charge a relatively high transaction fee that makes them infeasible for developers, given the small amount and volume of donations. Moreover, such platforms are not available to everyone because of country-specific legal restrictions or limited access to financial institutions. To overcome these challenges, developers started accepting cryptocurrencies as a payment method for receiving donations. Among many cryptocurrencies, Bitcoin is the most popular with a multi-billion dollar market.\footnote{At the time of writing, the market cap for Bitcoin is more than 107 billion dollars~\cite{coinmarketcap}.}

While Bitcoin and GitHub have co-existed for 10 years, we still do not know how prominent Bitcoin donations are, specifically to open source repositories, and how repository-specific factors, such as popularity metrics, affect the amount and volume of donations. In this paper, we bridge this knowledge gap and analyze the use of Bitcoin to make donations to open source repositories on GitHub.

We used Google BigQuery to scan over three million open source repositories looking for Bitcoin addresses, which are explicitly listed in ``readme'' files by repository owners for receiving donations (\S\ref{sec:dataset}). We then manually reviewed and validated the collected addresses, ending up with 1,996 donation addresses associated with 6,075 repositories. We also developed a web crawler and used it to directly fetch repository-specific information from GitHub, such as repository creation date, whether it is a fork, and popularity metrics like number of watchers, stars, and forks.

To analyze Bitcoin donations, we first extracted all transactions that involve any of the collected donation addresses from Bitcoin's public blockchain, ending up with 56,454 transactions. We then performed transaction and correlation analyses (\S\ref{sec:analysis}). Our results show that there is a limited adoption of Bitcoin as a payment method for receiving donations, and that the amount of donations does not strongly correlate with the popularity of a repository. We found that only 0.2\% of the repositories use Bitcoin for receiving donation, and that for those which do, only 51.8\% of their donation addresses have ever received deposits. In fact, just 9.2 thousand bitcoins have been donated to open source repositories on GitHub over the last 10 years, which is equivalent to 8.3 million dollars, calculated retrospectively. Moreover, the top-10 largest donations in value, which constituted 70.23\% of all donations in dollars, were made between late 2017 and early 2018 when the price of Bitcoin reached its record high. As such, this speculative and deflationary nature of bitcoin price could explain the disproportionately small amount of donations, where people are discouraged from making donations hoping for a higher future price. Another reason behind limited adoption of Bitcoin by repository owners is privacy, as once an address is publicly listed it could be linked to its owner and traced to personal transactions~\cite{boshmaf2019blocktag}.

We also found weak positive correlation between donation value in dollars and repository popularity, where the highest correlation ($r$=0.013) was associated with number of forks. Indeed, the top-10 grossing donation addresses, which received 54\% of all donations, are associated with unpopular repositories. In particular, six of these addresses are associated with single repositories that each had at most two watchers, stars, or forks. To some extent, this counter intuitive result can be explained by the fact that some unpopular repositories list personal addresses, which might receive deposits for reasons other than donations, or donation addresses of well-known open source organizations, which attract larger amounts of donations from channels other than GitHub.

To this end, this paper contributes with the first donation study which depends on actual data collected from public software repositories and cryptocurrency blockchains, instead of collecting self-reported data through user surveys. Even though the results draw a grim image, we remind the reader that people and institutions just started to seriously consider Bitcoin as a currency or an asset class, and that recent considerable donations, such as the one million dollar transaction to the Free Software Foundation by the anonymous Pineapple Fund~\cite{PineappleDonation}, are promising examples of how Bitcoin can be used to support the development of open source software. 

\section{Background}
\label{sec:background}

We now present a brief background on Bitcoin and GitHub. 

\subsection{Bitcoin}
\label{sec:background_bitcoin}

Bitcoin is the first and the most popular cryptocurrency network~\cite{nakamoto2008bitcoin}. In Bitcoin, the identity of a user is hidden by using public pseudonyms called addresses. A Bitcoin address is an alphanumeric identifier that is derived from the public key of a public/private key pair. It has three different formats in use today, each defining an address type, as specified by the regular expression patterns shown in Table~\ref{tab:bitcoin_regex}.

\begin{table}
\centering
\caption{Bitcoin address types and regex patterns.}
\label{tab:bitcoin_regex}    
\small
\begin{tabular}{llr}\toprule
    Type & Description & Regex pattern\\ \midrule
    P2PKH & Pay to public key hash & \texttt{1[a-km-zA-HJ-NP-Z1-9]\{25,34\}}\\
    P2SH& Pay to script hash & \texttt{3[a-km-zA-HJ-NP-Z1-9]\{25,34\}}\\
    Bech32 & Segregated witness & \texttt{bc1[a-zA-HJ-NP-Z0-9]\{25,39\}}\\
\bottomrule
\end{tabular}
\end{table}

The set of public/private keys that are owned by a user is called a wallet. Private keys are used to sign inputs of transactions as a proof of ownership. For example, if Alice wants to donate bitcoins to Bob, she creates a new transaction specifying one or more addresses from her wallet as inputs. She also specifies the amount to be transferred, as permitted by the amount previously received on the used inputs, and chooses one or more addresses from Bob's wallet as transaction outputs. To protect the transaction, she signs it using her private keys, and then broadcasts it to the whole network for verification.

All Bitcoin transactions are stored in a decentralized ledger called a blockchain, which also means anyone could try to identify user transactions by analyzing the blockchain. This task, however, is typically difficult since user identities are not recorded in the blockchain, only Bitcoin addresses. As such, Bitcoin does not provide complete anonymity but rather pseudonymity. If user identities are linked to Bitcoin addresses, their transactions can be easily identified. This is the case when users publicly post their Bitcoin addresses online along with their personally identifiable information, such as posting personal donation addresses on public GitHub repositories.

It is possible to exchange bitcoins with fiat currencies using cryptocurrency exchanges. While largely unregulated, the price is often based on supply and demand, but can still vary significantly during a day. Services like CoinMarketCap~\cite{coinmarketcap} and Yahoo Finance~\cite{YahooFinance} aggregate pricing data from exchanges and provide market statistics, such as open/close and high/low values for a trading day.

\subsection{GitHub}
\label{sec:background_github}

With more than 31 million developers worldwide~\cite{octoverse}, GitHub is one of the most popular web-based hosting services for Git version control. The service allows developers to upload their software projects into code repositories, which enables them to display, review, search, and navigate through their source code. Moreover, the service provides issue tracking, project planning, documentation, and release management tools. There is also a social networking feature that allows developers to watch, star, and fork repositories. As such, these social interactions can be used to estimate the popularity of a repository, as more popular repositories are likely to have a larger number of watchers, stars, and forks than others.

\begin{figure}
\centering
\includegraphics[width=0.8\columnwidth]{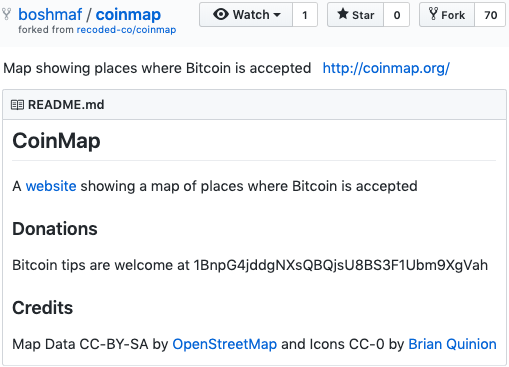}
\caption{Repository information on GitHub.}
\label{fig:github_repo}
\end{figure}

On GitHub, a user can create a new repository, either public or private, or fork (i.e., copy) an existing public repository. There are more than 96 million repositories hosted on GitHub, out of which 30 million are public and 3.4 million are open sourced under MIT, Apache, GPL, or other open source licenses~\cite{octoverse,GitHubActivityDataset}. Each repository has an identifier that consists of two parts: the repository's owner, which is the developer's alias on GitHub, and the repository's name. When forking a repository, the repository's name remains the same but the owner is changed to the forking developer. Figure~\ref{fig:github_repo} shows an example of how repository information is displayed on GitHub. From this figure, we know a repository called ``coinmap'' has been forked from a repository with the identifier ``recoded-co/coinmap'' by a developer with the alias ``boshmaf.'' 

We note that the displayed number of forks, also called network size, always refers to the root repository. For example, the number shown in Figure~\ref{fig:github_repo} means ``recoded-co'' had their repository forked 70 times, not ``boshmaf.'' However, using GitHub's API, it is possible to find out the number of immediate forks of a forked repository. To distinguish between these two numbers, we refer to the earlier as the network size and the latter as the number of forks.

In addition to repository information, GitHub also parses and displays text files called ``readme'' that are often used by developers to describe their work. By default, the service scans the ``root,'' ``docs'' and ``.github'' directories of a repository for such files~\cite{AboutGitHubReadmes}.

\section{Dataset}
\label{sec:dataset}

We now describe the dataset we used in this study.	

\subsection{Summary}
\label{sec:dataset_summary}

Overall, we collected 1,996 Bitcoin donation addresses associated with 6,075 open source repositories on GitHub. In terms of address types, P2PKH was the most popular with 1,843 addresses (92.33\%), followed by P2SH with 144 addresses (7.21\%), and finally the newly introduced Bech32 with 9 addresses (0.45\%).

As for the repositories, a total of 2,095 repositories (34.49\%) were forks. Overall, the repositories belonged to 4,686 owners, out of which 434 owners (9.26\%) represented organizations.

\subsection{Collection}
\label{sec:dataset_collection}

Our goal is to collect public information about open source repositories on GitHub that accept Bitcoin donations. As there are more than 96 million repositories hosted on GitHub~\cite{octoverse}, it is infeasible to download and inspect all of their contents. Instead, we used a recent snapshot of GitHub's open source repositories that is hosted on Google BigQuery service, which is available free of charge as part of Google Cloud Public Dataset Program~\cite{GoogleAnnouncement,GitHubAnnouncement}. We extracted a dataset of repositories that contain one or more Bitcoin addresses in their ``readme'' files. The snapshot we used in this study was updated on Dec 5, 2018, and includes files, commits, licences, and language information of over three million repositories~\cite{GitHubActivityDataset}.

We decided to search ``readme'' files because developers usually include a donation request as part of the overall repository description. In particular, we search ``root,'' ``docs,'' and ``.github'' directories of each repository for files whose name contains ``readme'' substring (e.g., ``README.md''). We then scan the content of each identified file for Bitcoin addresses using regular expressions, as specified in Table~\ref{tab:bitcoin_regex}. Google provides a possibility to search over the content of non-binary files on the ``HEAD'' branch that are less than 1MB~\cite{GoogleAnnouncement}. 

The snapshot we used does not include repository-specific information, such as its creation date, whether it is a fork or owned by an organization, and popularity-related information. To overcome this limitation, we developed a simple Python crawler to retrieve the missing information from Github, and download latest versions of ``readme'' files that contain Bitcoin addresses. We ran the crawler on Dec 26, 2018, and it took about 12 hours to update the dataset.

To this end, the collected dataset consisted of 7,968 repositories. Overall, there were 6,632 unique owners and 3,861 unique Bitcoin addresses, which means some owners have have several repositories, and some addresses are used in multiple repositories.

\subsection{Cleaning}
\label{sec:dataset_cleaning}

Some of the collected addresses might be invalid Bitcoin addresses. For example, an address that consists of all 1s or 3s is invalid. Moreover, even if an address is valid, is could be the case it is listed in the ``readme'' file for reasons others than accepting donations, such as showing a coding example that uses a well-known Bitcoin address.

In order to clean the dataset, we manually reviewed and validated the addresses associated with each repository against their use in ``readme'' files. We only kept addresses that are valid Bitcoin addresses which are used in a textual donation request, a donation link or badge, or the signature of a repository's owner. For example, Figure~\ref{fig:github_repo} shows an address that matches this criteria. Out of 3,861 addresses associated with 7,968 repositories, we ended up with 1,996 valid Bitcoin donation addresses associated with 6,075 repositories.

\section{Analysis}
\label{sec:analysis}

We now present the analysis we performed on the collected dataset.

\subsection{Summary}
\label{sec:analysis_summary}

Overall, we found a limited adoption of Bitcoin as a payment method for receiving donations, with nearly 44 thousand deposits adding up to only 8.3 million dollars in the last 10 years. We also found weak positive correlation between the amount of donation in dollars and the popularity of a repository, with highest correlation ($r$=0.013) associated with number of forks.

\subsection{Donations}
\label{sec:analysis_donations}

Our goal is to analyze the amount and volume of Bitcoin donations. To achieve this, we deployed BlockSci~\cite{Blocksci_Kalodner2017} on Jan 2, 2019, and used it to extract all transactions from Bitcoin's public blockchain which include one or more donation addresses as inputs or outputs, ending up with 56,454 transactions. If a transaction increases the balance of an address, we call it a deposit. Otherwise, we call the transaction a withdrawal. This classification is important when an address is used as both an input and an output of a transaction. We calculate the total deposit or withdrawal value for each donation address as the sum of increments or decrements of its balance in dollars, based on the exchange rate at the time of each corresponding transaction. As there is no evident way to exclude deposits to an address that are not donations, we refer to the total deposit value and the number of deposits as the amount and volume of donation, respectively. This means the results we report herein are optimistic and represent upper-bound estimates of the real amount and volume of donations.

Overall, a total of 8,348,241 dollars ($\approx$9,238.9516 bitcoins) were received through 43,862 deposits, and 25,889,923 dollars ($\approx$9,015.6384 bitcoins) were sent through 12,592 withdrawals. On average, a donation address received 4,182 dollars ($\approx$4.6287 bitcoins) through 21.97 deposits and sent 4,196 dollars ($\approx$4.5169 bitcoins) through 6.31 withdrawals. This money flow, and an average balance of 0.0958 bitcoins, show that deposits to donation addresses are rarely accumulated, but are rather spent or sent to other addresses as withdrawals.

In general, deposits and withdrawals are highly skewed across donation addresses. Only 1,034 addresses (51.8\%) have ever received deposits, with the top-10 grossing addresses receiving 70.23\% of all donations in dollars, or 37.44\% in bitcoins, as shown in Table~\ref{tab:top10_grossing_addresses}. Similarly, only 873 addresses (43.74\%) have ever been used in withdrawals. With an average repository age of 3.49 years, these findings suggest that Bitcoin has a limited adoption as a payment method for sending or receiving donations.\footnote{We could not measure the time at which a donation address is listed in a ``readme'' file, so we report the time at which the associated repository was created (i.e., its age).}

\begin{table*}
\centering
\caption{Top-10 grossing donation addresses in dollars.}
\label{tab:top10_grossing_addresses}
\small    
\begin{tabular}{lccrrrrrrrrr}
    \toprule
     & & & & \multicolumn{5}{c}{Repositories} & \multicolumn{3}{c}{Donations} \\
    \cmidrule(lr){5-9}
    \cmidrule(lr){10-12}
    & & & & Avg. age & \multicolumn{4}{c}{Total popularity} & \multicolumn{2}{c}{Amount} & Volume \\
    \cmidrule(lr){6-9}
    \cmidrule(lr){10-11}
    Address & Tag? & Org? & \# repos & (years) & \# watchers & \# stars & \# forks & Net. size & (bitcoins) & (dollars) & (txes) \\
    \midrule
        1GbAWBiGyu\ldots & \ding{55} & \ding{55} & 1 & 1.98 & 1 & 1 & 9 & 9 & 364.9295 & 1,689,358 & 113\\
        1257U991Wq\ldots & \ding{55} & \ding{55} & 1 & 2.55 & 1 & 0 & 0 & 0 & 1311.7105 & 1,302,104 & 144\\
        1PC9aZC4hN\ldots & \ding{51} & \ding{51} & 2 & 2.98 & 71 & 825 & 513 & 513 & 1085.1980 & 1,224,570 & 1772\\
        33ENWZ9RCY\ldots & \ding{51} & \ding{55} & 2 & 1.81 & 60 & 213 & 234 & 234 & 102.9535 & 365,765 & 25\\
        1MEWT2SGbq\ldots & \ding{55} & \ding{55} & 2 & 1.29 & 2 & 0 & 0 & 1602 & 80.7299 & 331,126 & 206\\
        1BsTwoMaX3\ldots & \ding{55} & \ding{55} & 1 & 3.29 & 1 & 0 & 0 & 151 & 284.1289 & 258,347 & 228\\
        1FFPahYGPH\ldots & \ding{55} & \ding{55} & 1 & 1.52 & 3 & 4 & 4 & 4 & 103.4162 & 220,196 & 95\\
        18xgGTTzZU\ldots & \ding{55} & \ding{55} & 1 & 3.80 & 1 & 0 & 0 & 0 & 46.9061 & 168,378 & 65\\
        3DKLj6rEZN\ldots & \ding{55} & \ding{51} & 1 & 2.59 & 22 & 158 & 11 & 11 & 23.4766 & 167,601 & 46\\
        1NV72LqZAJ\ldots & \ding{55} & \ding{55} & 1 & 2.16 & 43 & 211 & 59 & 59 & 55.6874 & 135,349 & 18\\
    \bottomrule
\end{tabular}
\end{table*}

\subsubsection{Top-grossing addresses}
\label{sec:analysis_donations_top_grossing}

While the average donation amount and volume per address is relatively low, a few addresses were clear outliers, as shown in Table~\ref{tab:top10_grossing_addresses}. In what follows, we manually inspect these addresses and the associated repositories looking for clues that could explain their success in receiving donations.

First, donation addresses associated with older and more popular repositories do not necessarily receive larger amounts of donations. While on average a donation address has 25.85 watchers, 281.87 stars, 75.85 forks, and a network size of 3,009.37 from nearly three repositories over 3.49 years, most of the addresses in Table~\ref{tab:top10_grossing_addresses} are associated with repositories that fall below these values. Also, all but one of the top-10 addresses which are listed in the most popular repositories, in terms of total number of stars, have received less than the average donation amount of 4,182 dollars, as shown in Table~\ref{tab:top10_popular_addresses}. We explore this relationship in more details in~\S\ref{sec:analysis_correlations}.

\begin{table}
\centering
\caption{Top-10 donation addresses in popular repositories.}
\label{tab:top10_popular_addresses}
\small   
\begin{tabular}{lrrrrr}
    \toprule
    & & & & \multicolumn{2}{c}{Donations}\\
    \cmidrule(lr){5-6}
    & & & Avg. age & Amount & Volume\\
    Address & \# repos & \# stars & (years) & (dollars) & (txes)\\
    \midrule
     1G8G6tqQ3o\ldots & 1 & 25,406 & 4.67 & 3,983 & 10\\
     1P9BRsmazN\ldots & 282 & 16,904 & 3.76 & 0 & 0\\
     1QDhxQ6Pra\ldots & 19 & 13,095 & 3.20 & 9,177 & 198\\
     1JnC15WwDV\ldots & 1 & 13,082 & 1.79 & 0 & 0\\
     3MDPzjXu2h\ldots & 1 & 11,315 & 4.55 & 9 & 1\\
     13PjuJcfVW\ldots & 10 & 11,057 & 2.55 & 0 & 0\\
     1EMqwwjqJr\ldots & 3 & 10,923 & 5.78 & 296 & 1\\
     1DGoNEYAnj\ldots & 9 & 10,371 & 3.25 & 29 & 8\\
     1MDmKC51ve\ldots & 1 & 10,212 & 5.37 & 96 & 6\\
     17NUKd3v7G\ldots & 17 & 10,167 & 3.59 & 5 & 1\\
    \bottomrule
\end{tabular}
\end{table}

Second, donation addresses can receive deposits from channels other than GitHub. In Table~\ref{tab:top10_grossing_addresses}, we mark an address as tagged if we find a public label describing it with auxiliary information from online services such as WalletExplorer~\cite{WalletExplorer}, Blockchain.Info~\cite{BlockchainInfo}, and BitcoinWhosWho~\cite{BitcoinWhosWho}. We also mark an address as organizational if any of its associated repositories is owned by a GitHub organization account. For example, the third address in Table~\ref{tab:top10_grossing_addresses} is tagged on Blockchain.Info as ``Free Software Foundation,'' and it turns out to be the foundation's official donation address~\cite{FSFDonations}. However, none of the associated repositories is owned by the foundation, which suggests that the donations were sent by individuals or organizations other than the users of these repositories. To support this claim, let us consider the top-10 largest deposits to donation addresses, as shown in Table~\ref{tab:top10_deposits_dollars}. The first deposit, with an amount of over one million dollars, was sent the foundation in early 2018, which is after the repositories were created. However, this deposit transaction is attributed to the anonymous Pineapple Fund, which donated a total of 55 million dollars (5,104 bitcoins) to various charities around the world between late 2017 and early 2018~\cite{PineappleDonation}.

\begin{table}
\centering
\caption{Top-10 deposits to donation addresses in dollars.}
\label{tab:top10_deposits_dollars}
\small    
\begin{tabular}{llrr}
    \toprule
    & & \multicolumn{2}{c}{Donations}\\
    \cmidrule(lr){3-4}
    & &  Amount & Received date\\
    Transaction & Address & (dollars) & (yyyy-mm-dd) \\
    \midrule
    145d80a087\ldots & 1PC9aZC4hN\ldots &  1,049,380 & 2018-01-29\\
    72f63ec8b6\ldots & 1GbAWBiGyu\ldots &  350,550 & 2017-12-17\\
    a0a1b2dd76\ldots & 18xgGTTzZU\ldots & 136,377 & 2017-08-13\\
    6c6af82ea3\ldots & 1GbAWBiGyu\ldots & 126,390 & 2017-09-11\\
    e869a0004f\ldots & 1GbAWBiGyu\ldots & 122,222 & 2017-09-01\\
    5281ed3ced\ldots & 1GbAWBiGyu\ldots & 105,248 & 2017-09-17\\
    a7bcbb803c\ldots & 1MEWT2SGbq\ldots & 102,852 & 2017-11-05\\
    c046f593f3\ldots & 1GbAWBiGyu\ldots & 82,567 & 2017-09-17\\
    14ee141b36\ldots & 33ENWZ9RCY\ldots & 77,153 & 2017-09-21\\
    050a02eadb\ldots & 1GbAWBiGyu\ldots & 72,039 & 2017-12-17\\
    \bottomrule
\end{tabular}
\end{table}

Third, donation addresses can receive deposits that are not necessarily donations. As shown in Table~\ref{tab:top10_grossing_addresses}, nearly all of the addresses are not tagged nor organizational, which means they are likely reused by their owners for various kinds of personal financial transactions. For example, the first address is associated with a repository that is owned by a Github user account from the European Union. The repository hosts PHP code for checking the status of a delegate node in the Shift network~\cite{Shift}. While the repository is unpopular and currently archived (i.e., read-only), the donation address has received six out of the top-10 largest amounts of deposits after the repository was created, constituting over 10\% of all donations in dollars, as shown in Table~\ref{tab:top10_deposits_dollars}. This suggests that such deposits are personal and not actual donations to the repository. Still, we cannot exclude them because there is no evident way to support this claim. 

\subsubsection{Historical perspective}
\label{sec:analysis_donation_historical}

\begin{figure}
\centering
\includegraphics[width=\columnwidth]{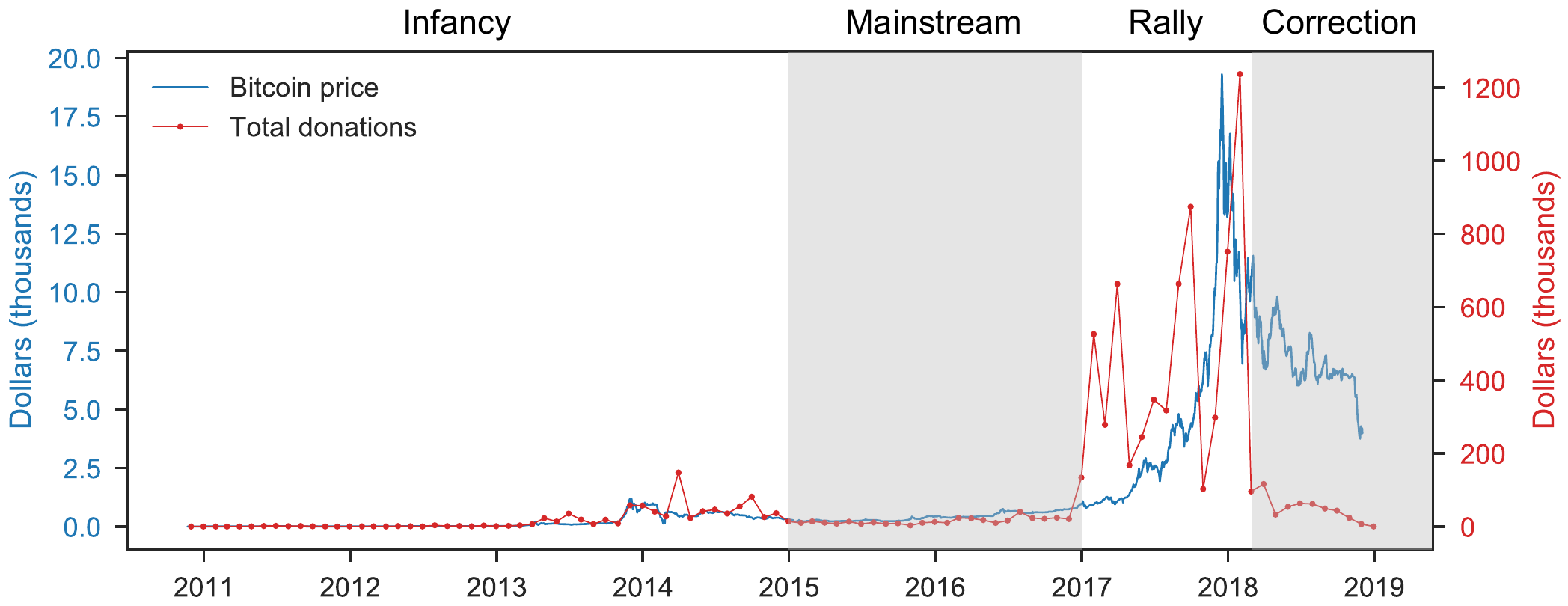}
\caption{Total amount of donations vs. exchange rate.}
\label{fig:price_vs_donations}
\end{figure}

\begin{figure}
\centering
\includegraphics[width=\columnwidth]{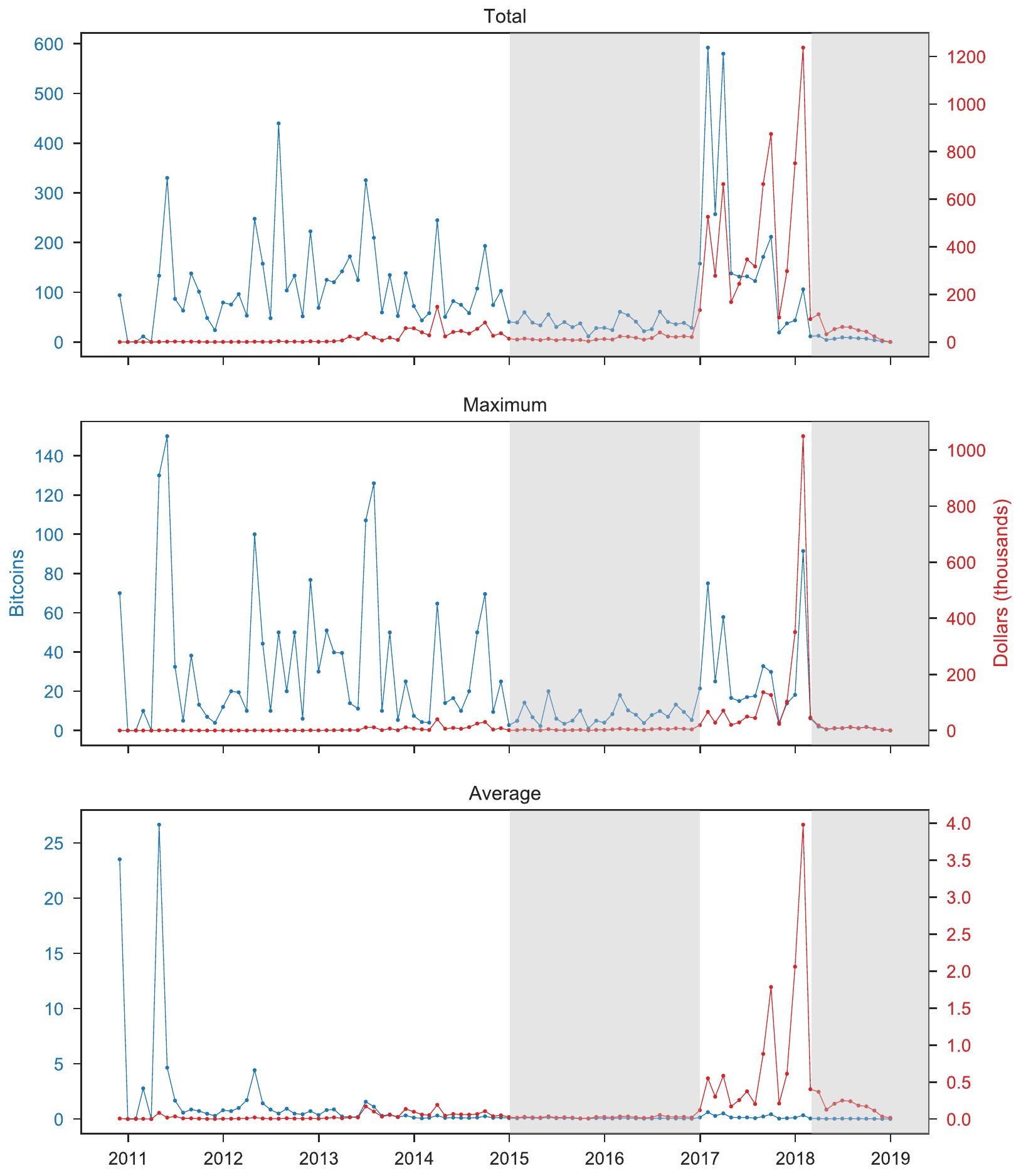}
\caption{Monthly amount of donations.}
\label{fig:monthly_donations}
\end{figure}

While it is reasonable to expect that the total amount of donations might increase as Bitcoin's popularity increases, we found that the exchange rate, which is often described as speculative, seem to have an impact as well. As shown in Figure~\ref{fig:price_vs_donations}, the total amount of monthly donations in dollars has increased in a relatively small increments until 2017, during which it has increased by orders of magnitude before plummeting down in 2018 onward. Although this change in value resembles the change in bitcoin price in dollars, the resemblance is unclear if we look at the total amount of donations in bitcoins instead of dollars, as shown in Figure~\ref{fig:monthly_donations}. To better understand this historical aspect of bitcoin donations, we next divide the last 10 years into four periods of time:

{\em Pre-2015 infancy.~}
Bitcoin was mostly used by early-adopters and much easier to mine, especially before 2013, when the mining reward was 50 bitcoins. Moreover, the cryptocurrency was difficult to buy or sell, as online exchanges were insecure and under-developed. As such, it is not surprising that people were comfortable giving away large amounts of bitcoins for donations. In fact, seven out of the top-10 largest deposits in bitcoins, which amount to \%8.29 of all donations, were received before 2015, as shown in Table~\ref{tab:top10_deposits_bitcoins}.

\begin{table}
\centering
\caption{Top-10 deposits to donation addresses in bitcoins.}
\label{tab:top10_deposits_bitcoins}
\small    
\begin{tabular}{llrr}
    \toprule
    & & \multicolumn{2}{c}{Donations}\\
    \cmidrule(lr){3-4}
    & &  Amount & Received date\\
    Transaction & Address & (bitcoins) & (yyyy-mm-dd) \\
    \midrule
    f9c2fc8da2\ldots & 1PC9aZC4hN\ldots &  150.0000 & 2011-05-07\\
    1d36a12fe7\ldots & 1PWC7PNHL1\ldots &  130.0000 & 2011-04-30\\
    30cbcfe01b\ldots & 1abrknajSF\ldots & 126.0000 & 2013-07-01\\
    6c6af82ea3\ldots & 1QCdQmJYUq\ldots & 107.0892 &2013-06-09\\
    1a6e430a62\ldots & 1PC9aZC4hN\ldots & 100.0000 & 2012-04-06 \\
    145d80a087\ldots & 1PC9aZC4hN\ldots & 91.4500 & 2018-01-29\\
    80ba4b1429\ldots & 1abrknajSF\ldots & 76.7460 & 2012-11-29\\
    0065017b4c\ldots & 1abrknajSF\ldots & 76.7460 & 2012-11-29\\
    b9fd76cd68\ldots & 1257U991Wq\ldots & 75.0000 & 2017-01-19\\
    abf03f3c25\ldots & 1DcZfySDvU\ldots & 70.0000 & 2010-11-27\\
    \bottomrule
\end{tabular}
\end{table}

{\em 2015--2016 mainstream.~} Bitcoin started to attract savvy investors, especially after its price reached that of an ounce of gold in early 2014. With a high expectation for higher returns and an increased liquidity from new exchanges, especially in China, people started to buy bitcoins and hold on them as gold. To some extent, this could explain the decrease in the total amount of donations in this period, which is mostly affected by the decrease in large deposits.

{\em 2017 rally.~} Bitcoin surged almost 400\% in a year. Thousands of ``Bitcoin millionaire'' gave rise to a wave of initial coin offerings and sent more than two billion dollars in funding to hundreds of new cryptocurrency projects~\cite{ForbesYearOfBitcoin}. This is also accompanied with large amounts of donations from mid 2017 to early 2018, where the top-10 largest deposits fall within this time period and constitute 26.65\% of all donations in dollars, as shown in see Table~\ref{tab:top10_deposits_dollars}.

{\em Post-2017 correction.~} Bitcoin went through a major price correction, losing more than 80\% of its value. People become more careful with their bitcoins, and the total amount of donations has plummeted to a few thousand dollars a month, without large deposits.

\subsection{Popularity and age}
\label{sec:analysis_correlations}

As shown in \S\ref{sec:analysis_donations}, some of the addresses that received large amounts of donations were listed in relatively new, unpopular repositories. Our goal is to further explore this relationship and analyze how repository information, its popularity and age in particular, are associated with the amount of donations it receives on listed address. In particular, we perform correlation analysis between five variables namely, the number of watchers, stars, and forks of a repository, its network size, and number of days since the repository was created, and the amount of received donations in bitcoins and dollars.

\begin{figure*}
\centering
\includegraphics[width=\linewidth]{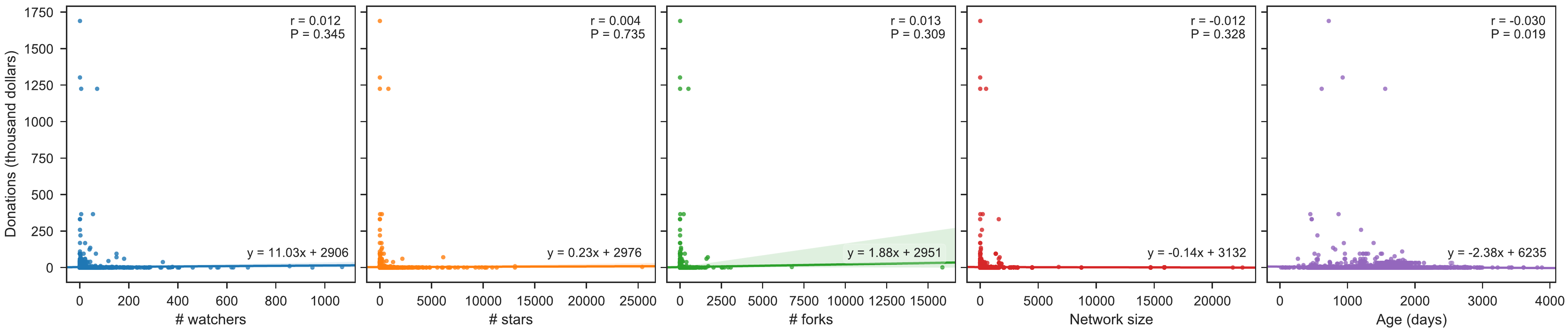}
\caption{Correlation between repository information and donations.}
\label{fig:repos_correlation_scatterplot}
\end{figure*}

We note that there are multiple addresses which are listed in one repository, and multiple repositories which list the same address. As repositories and addresses have this many-to-many relationship, we perform a separate analysis for each grouping and aggregation of the data, as follows.

\subsubsection{Repository}
\label{sec:analysis_correlations_repository}

\begin{figure}
\centering
\includegraphics[width=0.6\columnwidth]{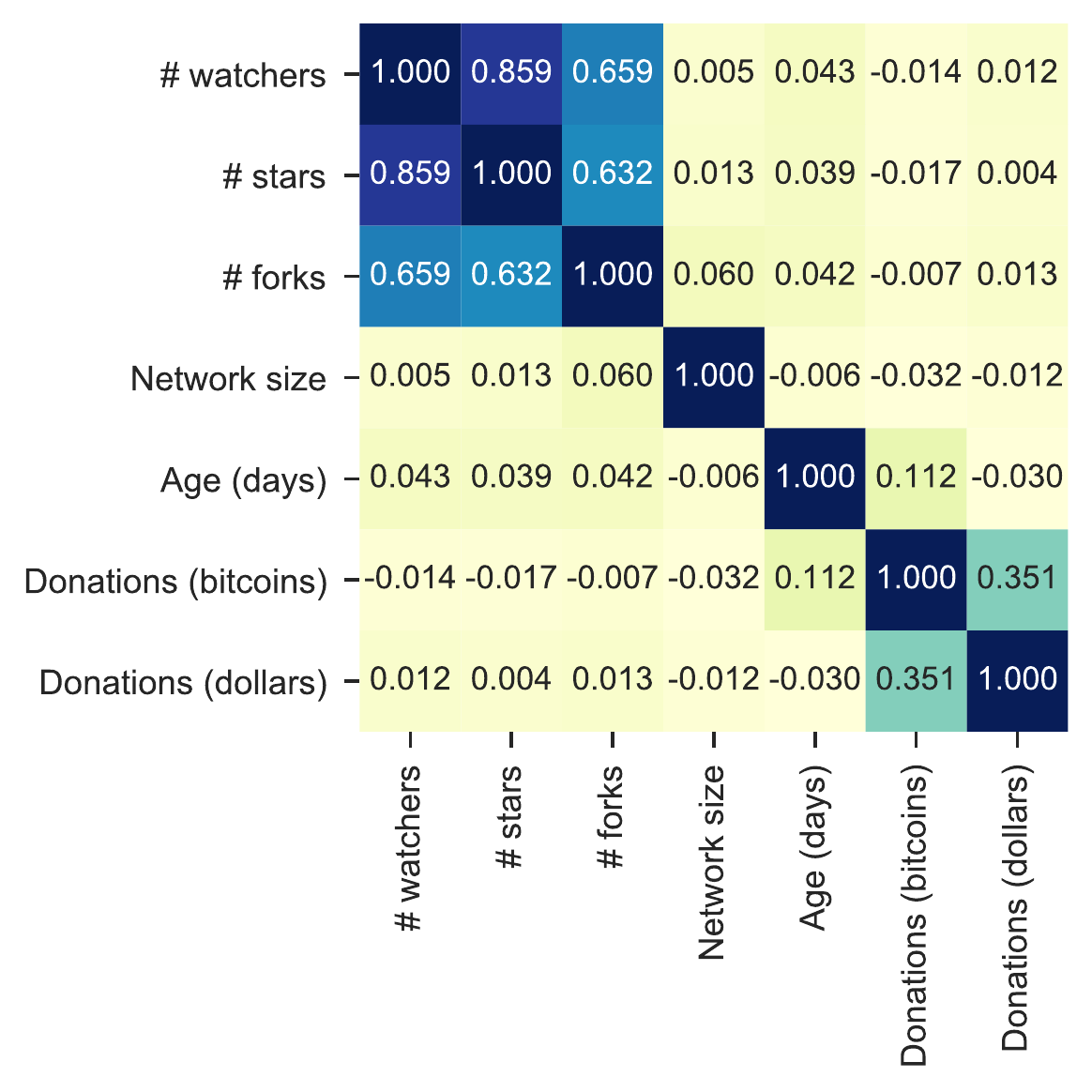}
\caption{Correlation matrix of repository information.}
\label{fig:repos_correlation_matrix}
\end{figure}

We grouped the data by each repository identifier, summing the amount of donations from all addresses listed in the repository as the total amount of donations. Figure~\ref{fig:repos_correlation_scatterplot} shows the correlation between repository information, represented as five different variables, and total amount of donations in dollars. We also plot the fit of a linear regression model for each variable with 95\% confidence interval. To evaluate pairwise correlation of all variables, including total amount of donations in dollars and bitcoins, we also computed the correlation matrix, as shown in Figure~\ref{fig:repos_correlation_matrix}. In what follows, we discuss the main findings from these two figures.

First, there is a weak positive correlation between number of watchers, stars, and forks and the total amount of donations in dollars, with the strongest correlation being the number of forks with $r$=0.013 and $P$=0.309. However, as shown in Figure~\ref{fig:repos_correlation_matrix}, there is a negative correlation between these variables and the total amount of donations in bitcoins, which is explained by the deflationary nature of bitcoin price compared to the dollar, even if the two amounts are positively correlated.

Second, from slope values of the fitted models, it is more important to have a larger number of watchers than forks, for instance, in relation to the total amount of donations. In fact, for each new watcher, star, or fork a repository attracts, a total of 11, 0.23, or 1.88 dollars in donations is expected, respectively. To some extent, this suggests that open source software projects could receive more donations if they attract more developers to contribute to their source code, not just users of their software, as software updates typically require forking the repository in order to send the new or modified code with the so called ``pull request.''

Third, while there is a weak positive correlation between the age of a repository and its popularity, there is a weak negative correlation between the age of a repository and the amount of donations in dollars. However, this correlation is positive when considering the amount of donations in bitcoins, which is expected, as older repositories did receive more deposits in bitcoins than new ones, when bitcoin's price was low and the cryptocurrency was not adopted by mainstream investors.

Fourth and last, the correlations should not be interpreted as statistically significant, as the associated P-values are relatively high, which also means that the amount of donations is considerably influenced by other factors. This can also be explained by the relatively high intercept values of the fitted models, which show that even unpopular repositories that do not have any watchers, stars, or forks, can receive a significant amount of donations on average---at least 2 thousand dollars in this case.

\subsubsection{Address}
\label{sec:analysis_correlations_address}

We shift the focus from repositories to addresses, and look at the relationship between address information, represented by aggregated information of repositories where they are listed, and the amount of donations they receive. The rationale behind this grouping is that some popular donation addresses that belong to a specific organization, such as the Free Software Foundation, are used in many repositories that build on top of their software, which could lead to different results.

\begin{figure*}
\centering
\includegraphics[width=0.575\linewidth]{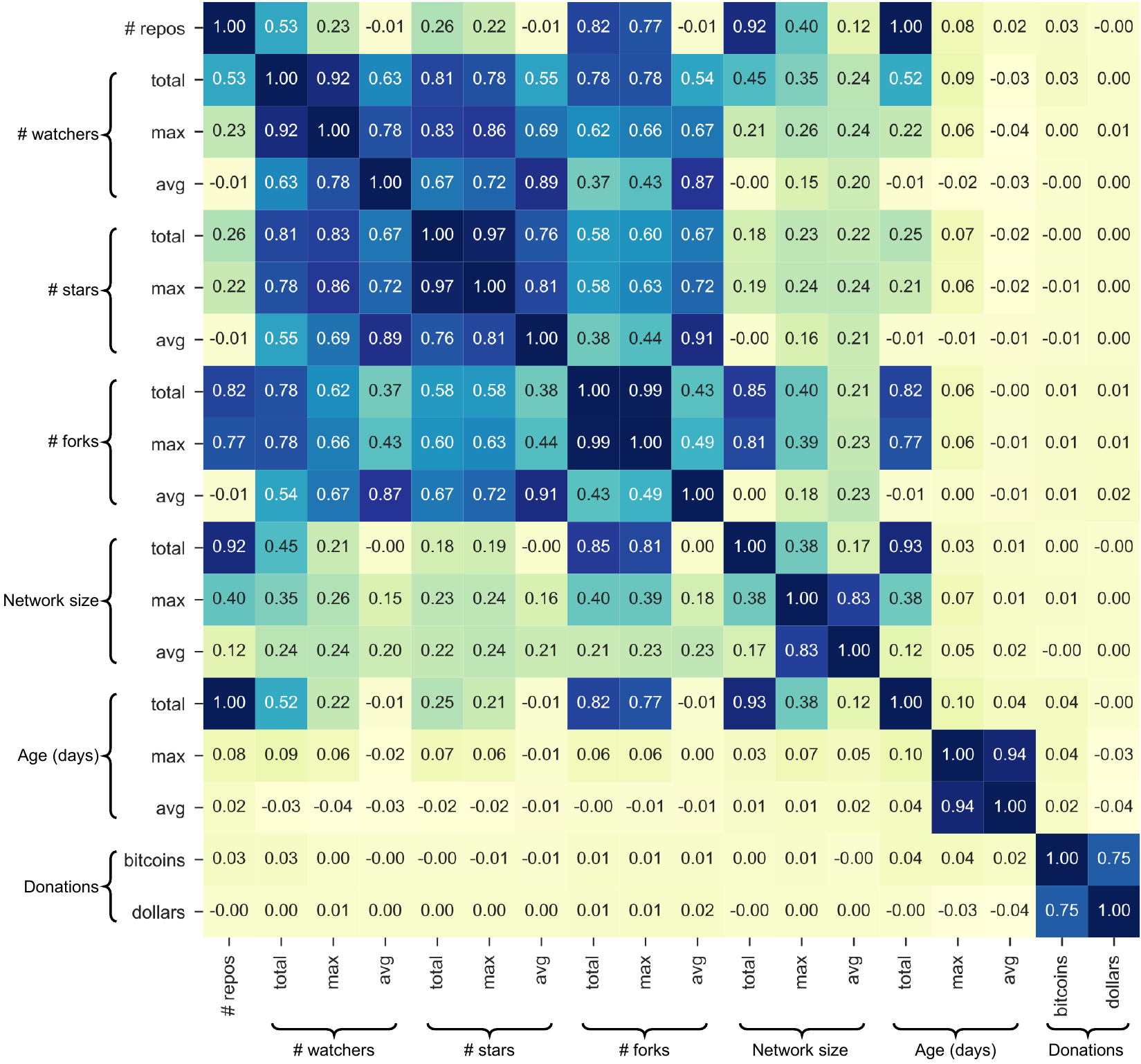}
\caption{Correlation matrix of address information.}
\label{fig:addresses_correlation_matrix}
\end{figure*}

As such, we grouped the data by each address, aggregating the total (i.e., sum), maximum, and average values per variable, in addition to calculating the number of repositories which list the address. However, as shown in Figure~\ref{fig:addresses_correlation_matrix}, we did not find strong correlation between address information and the amount of donations.

\section{Discussion}
\label{sec:discussion}

Bitcoin donations to open source repositories on GitHub are quite limited. On average, a repository receives 4,182 dollars worth of bitcoins in 3.49 years, which is around 100 dollars a month. The speculative and deflationary nature of bitcoin price could explain this disproportionately small amount of donations, as people tend to be discouraged from donating in bitcoins when they have expectations of proportionally higher future price. Another reason might be related to privacy, as once repository owners publicly list donation addresses they become vulnerable to deanonymization attacks, which aim to link their online identities (e.g., GitHub account) to Bitcoin transaction and activities with other users or services~\cite{boshmaf2019blocktag}. Regardless of the reason, even if it is a behavioral one, it is clear that with such a small amount of donations, repository owners and developers must rely on other sources of income and cannot fully dedicate their time to the hosted projects, unless they are being paid by their employers to work on them. Although open source software is often provided free of charge or not for profit~\cite{ghosh2002free}, larger amount of donations could surely help in improving its usability, quality, and security by attracting more full-time developers and using better infrastructures.

Repository information, such as popularity and age, do not seem to largely affect the amount of donations a repository is expected to receive. This counter intuitive result could be explained by the fact that some unpopular repositories list personal addresses, which receive deposits for reasons other than donations, or donation addresses of well-known open source organizations, which attract larger amounts of donations from channels other than GitHub.

\section{Limitations}
\label{sec:limitations}

The dataset is limited to licensed open source GitHub repositories, which represent about 10\% of all public repositories. While our focus is to analyze Bitcoin donations to open source software, considering all public repositories could lead to new insights about the problem. Moreover, Google BigQuery service hosts only text files on the ``HEAD'' branch that are less than 1MB in size, without access to file revisions, also called ``commit history.'' As such, we might have missed extracting some donation addresses because that information exists on a different branch, the address has been added after the snapshot was exported to BigQuery, or because ``readme'' files are larger than 1MB. While there are other web-based hosting services for open source repositories, such as GitLab~\cite{GitLab} and Bitbucket~\cite{Bitbucket}, we focus on GitHub as it is the most popular among developers, with more than 96 million repositories.

The analysis is limited to Bitcoin. While our manual address validation shows that developers ask for donations in other cryptocurrency networks, such as Ethereum and Litecoin, Bitcoin remains the most popular and widely-used network. We do not consider traditional, fee-based payment platforms, such as PayPal for one-off donations and Patreon for monthly donations, because they do not provide access to transaction data. While one could use different data collection methods, such as interviews and surveys, we focus on analyzing donations from evidence-based data sources, such as public blockchains, instead of having to rely on self-reported data.

\section{Related work}
\label{sec:related}

Donations are important for the development of open source software. In 2016, LibreOffice, a project of the Document Foundation, reported that they have received more than 200 thousand dollars in donations within a period of three years~\cite{libreoffice2016}. The foundation also emphasized that relying on a single sponsor is unnecessary as long as a diverse ecosystem of open source software community exists.

Studying donation economics of open source software is a challenging task. This is especially true for small-size software projects, as donations are typically anonymous and given directly to the people behind the project. Most of the existing research in this domain focuses on interviewing and surveying open source software developers to understand the motivations behind contributing to open source software~\cite{david2003floss, hars2001working, hertel2003motivation}. To the best of our knowledge, we are the first to analyze the amount and volume of donations from actual financial transactions stored in public blockchains.

DigitalOcean, a cloud infrastructure provider, has recently published as study on developer trends in the open source community~\cite{digitalocean2018}. They surveyed 4,300 full-time software developers who work at different companies about whether they use or contribute to open source software as part of their job. The results show that there is a disconnect between companies' encouragement of open source within their organizations, and their actual investment. While 71\% of the respondents reported that their companies expect them to routinely use open source software as part of their day-to-day development work, only 18\% of the respondents said their company is a member of an open source-related organization, and only 25\% said their company invests more than one thousand dollars every year in donations to open source software.

Ghosh et al.~\cite{ghosh2002free} studied the motivations behind contributing to open source software. They surveyed 2,784 developers from the open source software community and found that 4.4\% of the respondents would join a software project to make money and 12.3\% would stay involved with a project for the money. The authors conclude that developing open source software still resembles rather a hobby than salaried work.

While a large part of the Internet run on open source software, donations to the community, via Bitcoin or otherwise, are small. For example, the OpenSSL library serves 17.5\% of all web servers on the Internet, but the OpenSSL Software Foundation receives only two thousand dollars a year in donations~\cite{MoneyResponsibilityAndPride_Marquess2014}. This amount of money is not even enough to pay the developers behind the project, which makes external security audit financially infeasible, leading to infamous bugs like ``heartbleed''~\cite{HeartbleedBug_NetCraft2014}. Fortunately, after such large-scale security bugs were disclosed, large software companies started the initiative to fund open source software~\cite{InitiativeToFundOpenSSL_Arstechnica}.

\section{Future Work}
\label{sec:future_work}

We plan to address the limitation of this study in our future work. In particular, we are currently collecting a larger, more diverse dataset from all public repositories, not just licensed open source ones, from GitHub, GitLab, and Bitbucket. We are also working on enhancing the crawler to fetch the commit history of ``readme'' files. This will allow us to filter out transactions that were made before an address is listed in a repository, which could be after the repository was created. In addition, we can detect when an address is updated (i.e., changed or removed), and update the analysis accordingly. 

We plan to extend the study and consider more cryptocurrency networks, at least Ethereum, in order to get a comparative view of these networks as payment channels for donations. Finally, we plan to conduct a user study with repository owners in order to better understand the use of cryptocurrencies for donations, focusing on usability, security, and privacy issues.

\section{Conclusion}
\label{sec:conclusion}

We conducted a quantitative study analyzing Bitcoin donation to open source software on GitHub. Contrary to the previous research, which is mostly based on self-reported data, our study uses actual financial transactions stored in Bitcoin's public blockchain, which allowed us to calculate the amount and volume of donations over the last 10 years, and explore the relationship between the amount of donations received by a software repository and its hosting-related information, such as its age and popularity on GitHub.

Our results show that Bitcoin donations are unreasonably low, amounting to only 8.3 million dollars worth of bitcoins in the last 10 years, through 44 thousand deposits to six thousand repositories. We did not find strong correlations between the amount of donations and repository popularity or age, where the highest correlation ($r$=0.013, $P$=0.309) was associated with number of forks.

\section*{Acknowledgements}

We would like to thank the people at the Cybersecurity Initiative for Blockchain Research (CIBR) for their help and feedback.\footnote{For latest research outcomes, please visit \url{https://qcri.github.io/cibr}}

\bibliographystyle{ACM-Reference-Format}
\bibliography{refs}

\end{document}